\documentclass[namedreferences,hyperref,optionalrh,solaromanenum]{spr-sola}

\usepackage{graphicx}                    % For eps figures, newer & more powerfull
\usepackage{amsmath}                    % useful mathematical symbols
\usepackage{color}                       % For color text: \color command
\newcommand{\argmin}{\mathop{\mathrm{argmin}}}

\chardef\us=`\_

\newcommand{\vect}[1]{\boldsymbol{\mathbf{#1}}}
\newcommand{\torchmfbd}{\texttt{torchmfbd}}
\newcommand{\momfbd}{\texttt{MOMFBD}}

\DeclareMathOperator*{\argmax}{arg\,max}

\begin{document}

\begin{frontmatter}
\title{Marginal multi-object multi-frame blind deconvolution}

\author[addressref={aff1,aff2},email={e-mail.a@mail.com}]{\inits{}\fnm{Andr\'es}~\snm{Asensio Ramos}\orcid{0000-0002-1248-0553}}
%\author{\inits{}\fnm{}~\lnm{}\orcid{}}
%   NOTE:  Just one corresponding author [corref]
\address[id=aff1]{Instituto de Astrof\'isica de Canarias (IAC), Avda V\'ia L\'actea S/N,
    38200 La Laguna, Tenerife, Spain}
\address[id=aff2]{Departamento de Astrof\'isica, Universidad de La Laguna, 38205 La Laguna, Tenerife, Spain}

\runningauthor{Asensio Ramos}
\runningtitle{Marginal multi-object multi-frame blind deconvolution}

\begin{abstract}
    High-resolution ground-based solar imaging relies heavily on multi-object
    multi-frame blind deconvolution to correct for atmospheric turbulence.
    However, the traditional joint maximum likelihood estimation methods in which
    the object and the atmospheric aberrations are estimated together face
    some problems. In this paper, we introduce a marginal estimator for
    the multi-object multi-frame blind deconvolution problem. By employing a
    Bayesian framework to marginalize over the observed objects, we develop a
    stable reconstruction method that offers several distinct advantages over
    standard joint estimation. First, the marginalization provides enhanced
    regularization that naturally accounts for object uncertainty, successfully
    preventing the reconstruction algorithm from erroneously assigning noise
    power to high-order aberrations. Second, the marginal estimator yields more
    robust contrast control, as it is much less sensitive to the
    hyperparameters dictating the power spectral density (PSD) of the object
    prior. This robustness allows these hyperparameters to be optimized
    directly, enabling a ``plug-and-play'' deployment that removes the need for
    expert manual tuning. Finally, we demonstrate that the proposed method is
    highly accessible and simple to implement, requiring only the addition of a
    log-determinant term to the traditional merit function. With minimal
    modifications required for existing blind deconvolution pipelines, the
    marginal estimator has been fully integrated into the open-source \texttt{torchmfbd}
    package for its use by the solar physics community.
\end{abstract}
\end{frontmatter}

%
%-------------------------------------------------------------------
\section{Introduction}
The Earth’s atmosphere introduces refractive index fluctuations that severely degrade the
quality of solar observations. While ground-based solar telescopes are now almost
universally equipped with Adaptive Optics (AO) systems \citep{2020SoPh..295..172R,
2024A&A...685A..32S}, these systems only provide partial correction. Residual wavefront
errors persist, necessitating post-facto image restoration to reach the theoretical
diffraction limit of large-aperture telescopes.

Traditional restoration approaches generally fall into two categories: speckle
interferometry and model-based deconvolution. Speckle methods \citep{labeyrie70,vonderluhe93} rely on
the statistical properties of uncorrected turbulence, making them somehow difficult to apply to
AO-corrected data where the statistics are non-stationary. Speckle
methods are used in data from the GREGOR and DKIST \citep{2020SoPh..295..172R} telescopes. However, they 
require calibrations to produce correct outputs \citep{Fitzgerald_2006}.

Conversely, model-based methods like Multi-Object Multi-frame Blind Deconvolution \citep[MOMFBD;][]{2002SPIE.4792..146L,vannoort05} 
jointly estimate the solar object and the atmospheric aberrations
by maximizing the likelihood of the observed data. MOMFBD\footnote{We use \momfbd{} to refer to the code \citep{vannoort05,2021A&A...653A..68L} that is widely
  used for solar image restoration with the MOMFBD algorithm.} treats the restoration 
as a maximum likelihood (ML) problem, jointly estimating the solar object and the instantaneous wavefronts 
by maximizing the probability of the observed data under Gaussian or Poisson noise assumptions. MOMFBD
can handle a variety of data collection schemes. The main idea is acquiring
short-exposure images of different objects to produce improved images.
All objects (usually the same solar region observed, e.g., in
different wavelengths within a spectral line) have to be observed strictly
simultaneously, so that they are affected by exactly the same
turbulence. Additional phase diversity \citep[PD;][]{1982OptEn..21..829G,paxman92,paxman92phase} images 
can be acquired by introducing a known aberration (usually defocus) in the optical path, which
provides additional constraints to the problem and improves the quality of the restoration.
MOMFBD is routinely used with and without PD for the data from the CRisp Imaging Spectropolarimeter
\citep[CRISP;][]{2006A&A...447.1111S, 2008ApJ...689L..69S} and
CHROMospheric Imaging Spectrometer
\citep[CHROMIS;][]{2017psio.confE..85S} of the SST through its data processing 
pipeline SSTRED \citep{jaime15,2021A&A...653A..68L}.

Despite its success, the ML approach used in MOMFBD faces
some challenges in low signal-to-noise ratio (SNR) regimes. Because ML seeks a
joint point estimate for both the object and the wavefront (usually parameterized in terms of
an orthogonal decomposition of the pupil phase), the problem is often ill-posed and requires a
sufficiently large number of images. In the presence of noise, the algorithm tends
to overfit, assigning noise power to high-order aberrations. This requires a somehow
artisanal Fourier filtering in the final image to remove these artifacts and also 
a fine tuning of the regularization parameters.

In this paper, we propose a shift from joint point-estimation to a marginal maximum a-posteriori (MMAP)
framework, extending the work of \cite{blanco11} and \cite{fetick20} to the multi-object multi-frame
case. Instead of simultaneously solving for the object and the aberrations, we treat
the objects as "nuisance parameters." By defining a suitable Bayesian prior for them
and marginalizing (integrating) over the objects, we can isolate
the posterior probability of the aberrations alone. This marginalization naturally regularizes
the problem. It accounts for the uncertainty in the objects rather than
picking a single, potentially noisy point estimate.

By integrating over the distribution of possible objects, the resulting
restoration is less sensitive to high-frequency noise and can provide a more stable
reconstruction of multi-object, multi-frame data. We describe the mathematical formulation
of this marginal posterior method and show some of its properties. The necessary changes are 
minimal and can be easily implemented in existing codes. It has been already implemented in \torchmfbd\ \citep{2025A&A...703A.269A}.

\section{The inverse problem}
\subsection{Setup}
We consider an observation consisting of $J$ short-exposure frames for each of $K$
co-spatial objects with $L$ diversity channels. All $K$ objects of the $L$ diversity channels 
are observed strictly simultaneously, meaning they 
are all affected by the same atmospheric turbulence at any given instant.

If the exposure time for each frame is short enough (typically a few milliseconds), the 
atmospheric turbulence can be considered frozen during the exposure. If, additionally, 
the total duration of the burst of $J$ frames is shorter than the evolution
timescale of the observed scene, we can safely assume that the underlying objects remains
constant throughout the observation sequence. In a given isoplanatic patch, the
observed image $I_{kjp}$ at time $j=1,\ldots,J$ at the wavelength of object $k=1,\ldots,K$ and for the diversity channel $l=1,\ldots,L$
can be modeled as the convolution of the true object $o_k$ with a 
Point Spread Function (PSF) $P_{kjl}$ that represents the instantaneous atmospheric turbulence, plus an additive noise term $N_{kjl}$.
Following \cite{vannoort05}, we condense the time index $j$ and the diversity channel index $l$ into a single index $j$ 
that runs from 1 to $M=JL$, so that we can write the model as:
\begin{equation}
I_{kj} = P_{kj} * o_k + N_{kj},
\end{equation}
where $*$ denotes convolution, and the diversity is taken into account when necessary. 
Using the standard procedure, the PSFs are obtained from the autocorrelation of the generalized pupil function:
\begin{align}
    G_{kj}(\mathbf{r}) &= A(\mathbf{r}) e^{i \phi_{kj}(\mathbf{r})} \nonumber \\
    P_{kj} &= \left| \mathcal{F}^{-1}(G_{kj}) \right|^2,
\end{align}
with $A(\mathbf{r})$ being the aperture function as a function of the coordinates in the pupil plane
and $\phi_{kj}(\mathbf{r})$ the instantaneous wavefront phase, that is
expanded in Karhunen-Lo\`eve (KL) modes \cite[see][for details]{2025A&A...703A.269A}.

When images are digitized into $N \times N$ pixels, the convolution can be 
represented as a matrix-vector multiplication. 
By vectorizing the 
observed image $I_{kj}$ and the object $o_k$ into column vectors $\mathbf{I}_{kj}$ and $\mathbf{o}_k$ respectively, 
and representing the 
convolution PSFs as a Toeplitz matrix $\mathbf{H}_{kj}$, we can rewrite the model as:
\begin{equation}
    \mathbf{I}_{kj} = \mathbf{H}_{kj} \mathbf{o}_k + \mathbf{n}_{kj},
\end{equation}
where $\mathbf{n}_{kj}$ is the vectorized noise term. An additional level of abstraction can be obtained by concatenating all the 
observed images of a given object into a single vector $\mathbf{I}_k$:
\begin{align}
    \mathbf{I}_k &= [\mathbf{I}_{k1}^T, \mathbf{I}_{k2}^T, \dots, \mathbf{I}_{kM}^T]^T \nonumber \\
    \mathbf{H}_k &= [\mathbf{H}_{k1}^T, \mathbf{H}_{k2}^T, \dots, \mathbf{H}_{kM}^T]^T \nonumber \\
    \mathbf{n}_k &= [\mathbf{n}_{k1}^T, \mathbf{n}_{k2}^T, \dots, \mathbf{n}_{kM}^T]^T
\end{align}
so that we 
can write the model in a compact form as:
\begin{equation}
    \mathbf{I}_k = \mathbf{H}_k \mathbf{o}_k + \mathbf{n}_k,
\end{equation}
where $\mathbf{H}_k$ is a block-diagonal block-Toeplitz matrix containing the vertically stacked PSF matrices $\mathbf{H}_{kj}$ for 
each object.

\subsection{Bayesian formulation}
In a Bayesian framework, all the information is encoded on 
the posterior distribution of the unknowns. We make the assumption that the $K$ objects are statistically
independent, so that the posterior distribution factorizes as the product of the posterior distributions of each object.
For this reason, we focus our derivations on one specific object $k$ and drop the object
index to avoid cluttering the notation. The posterior distribution of the unknowns
(the object $\mathbf{o}$ and the PSFs $\mathbf{H}$) given the observed data $\mathbf{I}$,
which can be expressed using Bayes' theorem as:
\begin{equation}
    p(\mathbf{o}, \mathbf{H}, \vect{\beta}| \mathbf{I}) = \frac{p(\mathbf{I} | \mathbf{o}, \mathbf{H}, \vect{\beta}) 
    p(\mathbf{o}|\vect{\beta}) p(\mathbf{H}|\vect{\beta}) p(\vect{\beta})} {p(\mathbf{I})},
    \label{eq:bayes}
\end{equation}
where $p(\mathbf{I} | \mathbf{o}, \mathbf{H}, \vect{\beta})$ is the likelihood, discussed in the next section, and
$p(\mathbf{o}|\vect{\beta})$ and $p(\mathbf{H}|\vect{\beta})$ are the prior distribution of the object
and PSFs, respectively. Both the likelihood and priors are parameterized by the hyperparameters $\vect{\beta}$,
which can be used to control their properties.
Finally, $p(\vect{\beta})$ is the prior distributions of the hyperparameters
and $p(\mathbf{I})$ is the marginal likelihood (also known as evidence), which is a normalization constant, that
is unimportant when doing parameter estimation.

Although the posterior distribution contains all the information about the unknowns, it is often intractable 
to characterize it fully due to its high dimensionality and complex structure.
As a consequence, the problem is often solved by producing a point estimate, typically by 
finding the maximum a-posteriori (MAP) estimate, which is the mode of the posterior distribution.
We note that, from a numerical point of view, it is convenient to work with the negative 
log-posterior ($J=-\log p$), so that Eq. (\ref{eq:bayes}), can be rewritten as:
\begin{equation}
    J(\mathbf{o}, \mathbf{H}, \vect{\beta}|\mathbf{I}) = C + J(\mathbf{I}|\mathbf{o}, \mathbf{H}, \vect{\beta}) 
    + J(\mathbf{o}|\vect{\beta}) + J(\mathbf{H}|\vect{\beta}) + J(\vect{\beta}),
\end{equation}
where $C$ absorbs all constant terms independent of $\mathbf{o}$, $\mathbf{H}$, and $\vect{\beta}$.

\subsubsection{Likelihood}
Under the assumption of additive white Gaussian noise with zero mean and covariance matrix $\mathbf{\Sigma}$, the likelihood
follows is $p(\mathbf{I} | \mathbf{o}, \mathbf{H}, \vect{\beta}) = \mathcal{N}(\mathbf{I}| \mathbf{H} \mathbf{o}, \mathbf{\Sigma(\vect{\beta})})$,
a Gaussian distribution centered at the noise-free projection $\mathbf{H}\mathbf{o}$ and with 
covariance matrix $\mathbf{\Sigma(\vect{\beta})}$, which can potentially depend on the hyperparameters $\vect{\beta}$. 
Its explicit expression is:
\begin{equation}
    p(\mathbf{I} | \mathbf{o}, \mathbf{H}, \vect{\beta}) = (2\pi)^{-{P/2}} \det {\mathbf{\Sigma(\vect{\beta})}^{-1/2}} 
    \exp \left( -\frac{1}{2} (\mathbf{I} - \mathbf{H}\mathbf{o})^T \mathbf{\Sigma(\vect{\beta})}^{-1} (\mathbf{I} - \mathbf{H}\mathbf{o}) \right),
\end{equation}
where $P$ is the total number of pixels in all considered images and the covariance matrix is 
parameterized in terms of the hyperparameters $\vect{\beta}$. Since all images of a burst and 
objects are considered statistically independent, the likelihood factorizes as the product of the likelihoods of each image:
\begin{equation}
    p(\mathbf{I} | \mathbf{o}, \mathbf{H}, \vect{\beta}) = \prod_{k=1}^{K} \prod_{j=1}^{M} p(\mathbf{I}_{kj} | \mathbf{o}_k, \mathbf{H}_{kj}, \vect{\beta}).
\end{equation}

% \subsection{Likelihood}
% When you have $M$ different observations (e.g., multi-frame imaging or different sensors), we denote the 
% $i$-th observed image as $\mathbf{I}_i$. Each observation may have its own PSF $\mathbf{H}_i$ and noise characteristics $\mathbf{\Sigma}_i$.
% The total observation vector becomes a concatenation:
% \begin{equation}
%     \mathbf{I} = [\mathbf{I}_1^T, \mathbf{I}_2^T, \dots, \mathbf{I}_M^T]^T
% \end{equation}
% If the noise in each image is independent, the joint likelihood is the product of the individual likelihoods:
% \begin{equation}
%     p(\mathbf{I}_1, \dots, \mathbf{I}_M | \mathbf{o}, \mathbf{H}) = \prod_{i=1}^{M} p(\mathbf{I}_i | \mathbf{o}, \mathbf{H}_i)
% \end{equation}
% One can also write the likelihood as a single large multivariate Gaussian:
% \begin{equation}
%     p(\mathbf{I} | \mathbf{o}, \mathbf{H}) = \mathcal{N}(\mathbf{I}| \mathbf{H} \mathbf{o}, \mathbf{\Sigma}),
% \end{equation}
% where $\mathbf{H}$ is the vertically stacked PSF matrix $\mathbf{H} = [H_1^T, H_2^T, \dots, H_M^T]^T$ and
% $\mathbf{\Sigma}=\text{diag}(\Sigma_1, \Sigma_2, \dots, \Sigma_M)$ is the block-diagonal covariance matrix.

\subsubsection{Priors}
The choice of the priors $p(\mathbf{o}|\vect{\beta})$, $p(\mathbf{H}|\vect{\beta})$ and $p(\vect{\beta})$ is 
important for regularizing the problem and guiding the solution towards physically plausible results. Common choices 
include Gaussian priors for the object (which can encode assumptions about smoothness) or sparse priors (if images
are sparse in pixel space or any other domain). Priors on the 
PSFs should reflect our knowledge of atmospheric turbulence statistics. The hyperparameters can be assigned non-informative 
priors or can be treated as parameters to be optimized.

For simplicity, because it allows for analytical manipulation and also for the good results it
provides, it is customary to assume a Gaussian prior for the object:
\begin{equation}
    p(\mathbf{o}|\vect{\beta}) = \mathcal{N}(\mathbf{o}| \mathbf{m}, \mathbf{R}_o(\vect{\beta})),
\end{equation}
where $\mathbf{m}$ is the prior mean and $\mathbf{R}_o(\vect{\beta})$ is the prior covariance matrix.
We show later how the $\vect{\beta}$ hyperparameters can be used to control the shape of the prior covariance.
% The negative log-posterior is then given by:
% \begin{align}
% J(\mathbf{o}, \mathbf{H}, \vect{\beta}|\mathbf{I}) &= C + \frac{1}{2} \log \det \mathbf{\Sigma} + \frac{1}{2} \log \det \mathbf{R_o} \nonumber \\
% &+ \frac{1}{2} (\mathbf{I} - \mathbf{H} \mathbf{o})^T \mathbf{\Sigma}^{-1} (\mathbf{I} - \mathbf{H}\mathbf{o}) + 
% \frac{1}{2} (\mathbf{o} - \mathbf{m})^T \mathbf{R}_o^{-1} (\mathbf{o} - \mathbf{m}) \nonumber \\
% & + J(\mathbf{H}) + J(\vect{\beta}),
% \end{align}
% where $C$ absorbs all constant term.

\subsubsection{Posterior}
Since both the likelihood and the prior for the object are multivariate Gaussians, their product is 
itself a multivariate Gaussian distribution. Following \cite{2020arXiv200514199H}, this new Gaussian can be factorized as the following
product of two Gaussians:
\begin{equation}
    \mathcal{N}(\mathbf{I}| \mathbf{H}\mathbf{o}, \mathbf{\Sigma}) \mathcal{N}(\mathbf{o}| \mathbf{m}, \mathbf{R}_o) = 
    \mathcal{N}(\mathbf{o}| \mathbf{a}, \mathbf{A}) \mathcal{N}(\mathbf{I}| \mathbf{b}, \mathbf{B}),
    \label{eq:decomposition}
\end{equation}
where:
\begin{align}
    \mathbf{A}^{-1} &= \mathbf{H}^T {\mathbf{\Sigma}}^{-1} \mathbf{H} + {\mathbf{R}_o}^{-1}, \nonumber \\
    \mathbf{a} &= \mathbf{A} (\mathbf{H}^T \mathbf{\Sigma}^{-1} \mathbf{I} + \mathbf{R}_o^{-1} \mathbf{m}) \nonumber \\
    \mathbf{B} &= \mathbf{H} \mathbf{R}_o \mathbf{H}^T + \mathbf{\Sigma} \nonumber \\
    \mathbf{b} &= \mathbf{H} \mathbf{m}.
    \label{eq:decomposition_mean_covariance}
\end{align}
We point out that the factorization separates the dependence on the object $\mathbf{o}$ and the dependence on 
the observed data $\mathbf{I}$, which simplifies all calculations and is the key to an almost trivial
derivation of the joint and marginal estimators described in the following sections.

\subsection{Joint estimator}
The blind deconvolution problem has been classically solved as a maximum a-posteriori (MAP) estimate by
computing the object and PSFs that maximize the posterior:
\begin{align}
    \mathbf{o}_\mathrm{MAP}, \mathbf{H}_\mathrm{MAP} &= \argmax_{\mathbf{o}, \mathbf{H}} p(\mathbf{o}, \mathbf{H}, \vect{\beta} | \mathbf{I}) \nonumber \\
    &=\argmax_{\mathbf{o}, \mathbf{H}} p(\mathbf{I}|\mathbf{o}, \mathbf{H}, \vect{\beta}) 
    p(\mathbf{o}|\vect{\beta}) p(\mathbf{H}|\vect{\beta}) p(\vect{\beta}).
\end{align}
In the joint estimator, the hyperparameters $\vect{\beta}$ are fixed in advance and not optimized.
Equivalently, transforming to negative log-posteriors, the problem to solve is:
\begin{align}
    \mathbf{o}_\mathrm{MAP}, \mathbf{H}_\mathrm{MAP} &= \argmin_{\mathbf{o}, \mathbf{H}} J(\mathbf{o}, \mathbf{H}, \vect{\beta}|\mathbf{I}) \nonumber \\
    &= \argmin_{\mathbf{o}, \mathbf{H}} J(\mathbf{I}|\mathbf{o}, \mathbf{H}, \vect{\beta}) +
    J(\mathbf{o}|\vect{\beta}) + J(\mathbf{H}|\vect{\beta}) + J(\vect{\beta}).
\end{align}

Since $\mathcal{N}(\mathbf{I}| \mathbf{b}, \mathbf{B})$ in Eq. (\ref{eq:decomposition}) does not explicitly depend on the object, the 
maximum a-posteriori solution object fulfills $\mathbf{o}_\mathrm{MAP}=\mathbf{a}$, i.e., it is simply the mean of the 
first Gaussian of the right-hand side of Eq. (\ref{eq:decomposition}). Using the definition of $\mathbf{a}$ in 
Eq. (\ref{eq:decomposition_mean_covariance}), we find:
\begin{equation}
    \mathbf{o}_\mathrm{MAP} = (\mathbf{H}^T \mathbf{\Sigma}^{-1} \mathbf{H} + \mathbf{R}_o^{-1})^{-1} (\mathbf{H}^T \mathbf{\Sigma}^{-1} \mathbf{I} + \mathbf{R}_o^{-1} \mathbf{m}).
\end{equation}
This expression is a generalization 
of the classical Wiener filter used in the context of image restoration and deconvolution
\citep{1982OptEn..21..829G,paxman92,paxman92phase,lofdahl_scharmer94,2002SPIE.4792..146L,vannoort05}
to the case of a non-zero prior mean and a non-diagonal prior covariance. The comparison with these
previous works will be more clear when evaluated in the Fourier space as shown in Sec. \ref{sec:fourier}.

When plugging $\mathbf{o}=\mathbf{o}_\mathrm{MAP}$ back in Eq. (\ref{eq:decomposition}), the first Gaussian of 
the right-hand side becomes simply the normalization constant because the exponential is evaluated at 
its mean. The negative log-posterior can then be expressed as:
\begin{align}
J_\mathrm{joint}(\mathbf{H},\vect{\beta}|\mathbf{I},\mathbf{o}=\mathbf{o}_\mathrm{MAP}) &= C + \frac{1}{2} \log \det \mathbf{A} + \frac{1}{2} \log \det \mathbf{B}\nonumber \\
&+ \frac{1}{2} (\mathbf{I} - \mathbf{H} \mathbf{m})^T \mathbf{\mathbf{B}}^{-1} (\mathbf{I} - \mathbf{H} \mathbf{m}) \nonumber \\
& + J(\mathbf{H}|\vect{\beta}) + J(\vect{\beta}).
\label{eq:J_AB}
\end{align}
For clarity, we have explicitly used the label "joint" to refer to the fact that this is the negative log-posterior 
of the joint estimator.
Using the matrix determinant lemma and the fact that $\log \det \mathbf{A}^{-1}=-\log \det \mathbf{A}$, we can express 
$\log \det \mathbf{A}$ as:
\begin{equation}
    \log \det \mathbf{A} = \log \det \mathbf{R}_o + \log \det \mathbf{\Sigma} - \log \det \mathbf{B}
\end{equation}
so that Eq. (\ref{eq:J_AB}) simplifies to:
\begin{align}
J_\mathrm{joint}(\mathbf{H},\vect{\beta}|\mathbf{I}) &= C + \frac{1}{2} \log \det \mathbf{\Sigma} + \frac{1}{2} \log \det \mathbf{R}_o \nonumber \\
&+\frac{1}{2} (\mathbf{I} - \mathbf{H} \mathbf{m})^T \mathbf{\mathbf{B}}^{-1} (\mathbf{I} - \mathbf{H} \mathbf{m}) \nonumber \\
& + J(\mathbf{H}|\vect{\beta}) + J(\vect{\beta}),
\label{eq:log_joint_matrix}
\end{align}
where we dropped the condition $\mathbf{o}=\mathbf{o}_\mathrm{MAP}$ to avoid cluttering the expression.

% This very same expression has been used in the original formulation of the MOMFBD algorithm \citep{2002SPIE.4792..146L} and in 
% its implementation \momfbd{} \citep{vannoort05}, where 

\subsection{Marginal likelihood}
\label{sec:marginal}
The factorization of Eq. (\ref{eq:decomposition}) allows for a straightforward computation of the marginalized 
likelihood, which is the key to the marginal estimator. Instead of solving for the MAP object and plugging it
back in the posterior, we can integrate out the object from the joint posterior. This is equivalent to treating 
the object as a ``nuisance parameter'' and focusing on the posterior distribution of the PSFs alone.
Mathematically, this is defined as:
\begin{align}
    p(\mathbf{H}, \vect{\beta}|\mathbf{I}) &= \int p(\mathbf{o},\mathbf{H}, \vect{\beta}|\mathbf{I}) \, \mathrm{d}\mathbf{o} \nonumber \\
    &\propto 
    p(\mathbf{H}|\vect{\beta}) p(\vect{\beta})
    \int p(\mathbf{I} | \mathbf{o}, \mathbf{H}, \vect{\beta}) p(\mathbf{o}|\vect{\beta}) \, \mathrm{d}\mathbf{o}.
\end{align}
This integral can be trivially computed using the decomposition of Eq. (\ref{eq:decomposition}). Since
the Gaussians are normalized to unit area, the marginalized likelihood is simply
given by the second Gaussian:
\begin{equation}
    p(\mathbf{H}, \vect{\beta}|\mathbf{I}) = N(\mathbf{I}| \mathbf{b}, \mathbf{B}) p(\mathbf{H}|\vect{\beta}) p(\vect{\beta}).
\end{equation}
where $\mathbf{b}$ and $\mathbf{B}$ are defined in Eq. (\ref{eq:decomposition_mean_covariance}).
As a consequence, the negative log marginal likelihood is then given by:
\begin{align}
J_\mathrm{marg}(\mathbf{H},\vect{\beta}|\mathbf{I}) &= C + \frac{1}{2}\log \det \mathbf{B} \nonumber \\
&+ \frac{1}{2}(\mathbf{I} - \mathbf{H} \mathbf{m})^T \mathbf{B}^{-1} (\mathbf{I} - \mathbf{H}\mathbf{m}) \nonumber \\
& + J(\mathbf{H}|\vect{\beta}) + J(\vect{\beta}).
\label{eq:log_marginal}
\end{align}
Since all images share the same object $\mathbf{o}$, the covariance 
$\mathbf{B}$ will have off-diagonal blocks representing the correlation between different images 
induced by the shared prior. 

Both the joint and marginal metrics are very similar, and can be related by the following expression:
\begin{align}
    J_\mathrm{marg}(\mathbf{H},\vect{\beta}|\mathbf{I}) &= J_\mathrm{joint}(\mathbf{H},\vect{\beta}|\mathbf{I}, \mathbf{o}=\mathbf{o}_\mathrm{MAP}) \nonumber \\
    &- \frac{1}{2} \log \det \mathbf{R}_o
    - \frac{1}{2} \log \det \mathbf{\Sigma} + \frac{1}{2} \log \det \mathbf{B}.
    \label{eq:log_marginal_matrix}
\end{align}

\subsection{Efficient computation in the Fourier domain}
\label{sec:fourier}
In the spatial domain, the matrices $\mathbf{\Sigma}$, $\mathbf{H}$ and $\mathbf{R}_o$ are massive, making the 
inversion of $\mathbf{B}$ and the computation of its determinant computationally impractical. Fortunately, if we 
assume periodic boundary conditions in the images, 
the PSF matrices $\mathbf{H}_i$ and the prior covariance $\mathbf{R}_o$ become block-circulant with circulant block,
which are diagonalized by the 2D Discrete Fourier Transform matrix ($\mathbf{F}$):
\begin{align}
    \mathbf{H}_i &= \mathbf{F} \mathbf{\Lambda}_i \mathbf{F}^\dagger \\
    \mathbf{R}_o(\vect{\beta}) &= \mathbf{F} S_o(\vect{\beta}) \mathbf{F}^\dagger,
\end{align}
where $\mathbf{F}$ is the unitary 2D Fourier matrix ($\mathbf{F} \mathbf{F}^\dagger = \mathbf{1}$, with
$\mathbf{F}^\dagger$ being the conjugate transpose of $\mathbf{F}$ and $\mathbf{1}$ the identity matrix), 
The $\mathbf{\Lambda}_i$ matrices are diagonal with the Fourier components of the PSF on the diagonal,
and $S_o(\vect{\beta})$ is the power spectral 
density (PSD) of the object prior, parameterized with the hyperparameters $\vect{\beta}$. With this notation, the Fourier transform of the images and 
prior mean are given by $\tilde{\mathbf{I}}_i = \mathbf{F} \mathbf{I}_i$ and $\tilde{\mathbf{m}} = \mathbf{F} \mathbf{m}$.
The covariance $\mathbf{B}$ becomes much simpler in the Fourier domain:
\begin{equation}
\mathbf{B} = \mathbf{\Sigma} +
\begin{bmatrix} 
    \mathbf{\Lambda}_1 S_o \mathbf{\Lambda}_1^\dagger & \dots & \mathbf{\Lambda}_1 S_o \mathbf{\Lambda}_M^\dagger \\ 
    \vdots & \ddots & \vdots \\ 
    \mathbf{\Lambda}_M S_o \mathbf{\Lambda}_1^\dagger & \dots & \mathbf{\Lambda}_M S_o \mathbf{\Lambda}_M^\dagger
\end{bmatrix},
\end{equation}
with the second matrix being a block-diagonal matrix. For simplicity, we assume from now on that 
all images have uncorrelated Gaussian noise with the same variance $\sigma^2$, so that the covariance 
$\mathbf{\Sigma} = \sigma^2 \mathbf{1}$. This is an assumption that is almost always fulfilled
because the $M$ images are obtained in very rapid succession.
As a consequence, in the Fourier domain, every single one of the $N^2$ frequency components $u$ (e.g., pixel in the 
Fourier domain) is independent of the others. 
Therefore, we can decompose the matrix $\mathbf{B}$ as a block-diagonal matrix with $N^2$ blocks, each of which has size $M \times M$.
For each frequency index $u$, we let $\mathbf{y}_u = [\tilde{{I}}_1(u), \dots, \tilde{{I}}_M(u)]^T$,
$\mathbf{h}_u = [{\Lambda}_1(u), \dots, {\Lambda}_M(u)]^T$ 
and $s_u = S_o(u)$.
With this notation, the quadratic form that appears in Eqs. (\ref{eq:log_joint_matrix}) and (\ref{eq:log_marginal}) 
and the log-determinant term in Eq. (\ref{eq:log_marginal}) can be rewritten as a sum over frequencies:
\begin{align}
(\mathbf{I} - \mathbf{H} \mathbf{m})^T \mathbf{B}^{-1} (\mathbf{I} - \mathbf{H}\mathbf{m}) &=
\sum_{u}  (\mathbf{y}_u - \mathbf{h}_u \tilde{\mathbf{m}}_u)^\dagger \mathbf{Q}_u^{-1} (\mathbf{y}_u - \mathbf{h}_u \tilde{\mathbf{m}}_u) \nonumber \\
\log \det \mathbf{B} &= \sum_{u} \log \det \mathbf{Q}_u,
\label{eq:marginal_fourier}
\end{align}
where the sum over $u$ extends to all $N^2$ frequencies, and the $M \times M$ matrix $\mathbf{Q}_u$ is given by:
\begin{equation}
    \mathbf{Q}_u = \sigma^2 \mathbf{1} + s_u \mathbf{h}_u \mathbf{h}_u^\dagger.
\end{equation}
The inverse of the matrix $\mathbf{Q}_u$ and its determinant can be computed efficiently using the Woodbury matrix identity
(or the simpler Sherman–Morrison formula) and the determinant lemma. 
\begin{equation}
    \mathbf{Q}_u^{-1} = \frac{1}{\sigma^2}
    \left( \mathbf{1} - \frac{s_u \mathbf{h}_u \mathbf{h}_u^\dagger}{\sigma^2 + s_u \|\mathbf{h}_u\|^2} \right),
    \label{eq:qu_inverse}
\end{equation}
and
\begin{equation}
    \log \det \mathbf{Q}_u = (M-1)\log \sigma^2 + \log(\sigma^2 + s_u \|\mathbf{h}_u\|^2).
    \label{eq:log_det_qu}
\end{equation}

Substituting Eqs. (\ref{eq:qu_inverse}) and (\ref{eq:log_det_qu}) in Eq. (\ref{eq:marginal_fourier}), expanding the
quadratic term, and making explicit the summations over the images, Eqs. (\ref{eq:log_joint_matrix}) 
and (\ref{eq:log_marginal_matrix}) become:
% \begin{align}
% J_\mathrm{joint}(\mathbf{H},\vect{\beta}|\mathbf{I}) &= C + \frac{1}{2\sigma^2}
% \sum_{u=1}^{N^2} \left[ \|\mathbf{d}_u\|^2 - 
% \frac{s_u |\mathbf{h}_u^\dagger \mathbf{d}_u|^2}{\sigma^2 + s_u \|\mathbf{h}_u\|^2} \right],
% \label{eq:log_joint_fourier}
% \end{align}
% and
% \begin{align}
% J_\mathrm{marg}(\mathbf{H},\vect{\beta}|\mathbf{I}) &= C + \frac{1}{2} \sum_{u=1}^{N^2} 
% \log(\sigma^2 + s_u \|\mathbf{h}_u\|^2) + (M-1) \log \sigma^2 \nonumber \\
% &+ \frac{1}{2\sigma^2} \sum_{u=1}^{N^2} \left[ \|\mathbf{d}_u\|^2 - 
% \frac{s_u |\mathbf{h}_u^\dagger \mathbf{d}_u|^2}{\sigma^2 + s_u \|\mathbf{h}_u\|^2} \right],
% \end{align}
\begin{align}
J_\mathrm{joint}(\mathbf{H},\vect{\beta}|\mathbf{I}) &= C 
+\frac{1}{2\sigma^2} \sum_{u=1}^{N^2} \left[ \sum_{i=1}^{M} |\mathbf{d}_i(u)|^2 - 
\frac{s_u |\sum_{i=1}^{M} \mathbf{\Lambda}_i^*(u) \mathbf{d}_i(u)|^2}{\sigma^2 + s_u \sum_{i=1}^{M} |\mathbf{\Lambda}_i(u)|^2} \right],
\label{eq:log_joint_fourier}
\end{align}
and
\begin{align}
J_\mathrm{marg}(\mathbf{H},\vect{\beta}|\mathbf{I}) &= C + \frac{1}{2} \sum_{u=1}^{N^2} 
\log \left (\sigma^2 + s_u \sum_{i=1}^{M} |\mathbf{\Lambda}_i(u)|^2 \right) + (M-1) \log \sigma^2 \nonumber \\
&+\frac{1}{2\sigma^2} \sum_{u=1}^{N^2} \left[ \sum_{i=1}^{M} |\mathbf{d}_i(u)|^2 - 
\frac{s_u |\sum_{i=1}^{M} \mathbf{\Lambda}_i^*(u) \mathbf{d}_i(u)|^2}{\sigma^2 + s_u \sum_{i=1}^{M} |\mathbf{\Lambda}_i(u)|^2} \right].
\end{align}
where $\mathbf{d}_u = \mathbf{y}_u - \mathbf{h}_u \tilde{\mathbf{m}}_u$ is the residual vector at frequency $u$
and $C$ absorbs all constant terms independent of the PSFs and hyperparameters. The marginal estimator turns out
to be very similar, with the addition of the log-determinant terms. A final simplification occurs when the prior mean 
is zero ($\mathbf{m} = 0$), leading to $\mathbf{d}_u = \mathbf{y}_u$. We point out that Eq. (\ref{eq:log_joint_fourier})
with $\mathbf{d}_u = \mathbf{y}_u$ is the standard merit function used in \momfbd\ \citep{lofdahl_scharmer94,vannoort05}.
We point out that the DC component is usually not included in the summation over frequencies in the loss function. 
The reason is that the PSFs are always normalized to unit area, so that the DC component of the OTFs is always 1, leading
to a constant contribution to the loss function that does not depend on the PSFs. 

Finally, note that the MAP estimate of the object can also be computed efficiently in the Fourier domain. Using 
the diagonalization of $\mathbf{H}$ and
$\mathbf{R}_o$, we find that the Fourier coefficients of the MAP object estimate are given by:
\begin{equation}
    \tilde{\mathbf{o}}_\mathrm{MAP}(u) = \frac{s_u \sum_{i=1}^{M} \mathbf{\Lambda}_i^*(u) \tilde{\mathbf{I}}_i(u) + 
    \sigma^2 \tilde{m}_u}{\sigma^2 + s_u \sum_{i=1}^{M} |\mathbf{\Lambda}_i(u)|^2}.
    \label{eq:MAP_object}
\end{equation}
In practical terms, the MAP object estimate is found to be much more robust if it is computed 
using:
\begin{equation}
    \tilde{\mathbf{o}}_\mathrm{MAP}(u) = H \frac{\sum_{i=1}^{M} \mathbf{\Lambda}_i^*(u) \tilde{\mathbf{I}}_i(u)}
    {\sum_{i=1}^{M} |\mathbf{\Lambda}_i(u)|^2}.
    \label{eq:MAP_object2}
\end{equation}
with $H$ being a low-pass filter that is obtained following the same procedure as described in \cite{lofdahl_scharmer94}.

% \begin{figure}
%     \centerline{\includegraphics[width=0.8\textwidth]{figs/hyperpriors.pdf}}
%     \caption{Hyperpriors for $K$ and $u_0$. We fix $\mu_K=1$, $\mu_u=0.1$, and $\mu_p=2$, while
%     $\sigma_K=0.3$, $\sigma_u=0.5$, and $\sigma_p=2$.\label{fig:hyperpriors}}
% \end{figure}

\subsection{PSD parameterization}
The previous formalism requires the definition of several hy\-per\-pa\-ram\-e\-ters. In our case, they are 
the noise variance $\sigma^2$ and the parameters that control the shape of the
PSD of the object prior: $\vect{\beta} = \{\kappa, u_0, p, \sigma^2\}$. Following standard practice, we model the PSD using 
the following functional form:
\begin{equation}
    s_u(\vect{\beta}) = \frac{\kappa N}{\left[1 + \left(u/u_0 \right)^2 \right]^{p/2}},
\end{equation}
which behaves as a constant $K$ for low frequencies ($u \ll u_0$) and decays as a power law with 
exponent $p$ for high frequencies ($u \gg u_0$). The parameter $u_0$ controls the transition between 
the low-frequency and high-frequency regimes, while $p$ controls the steepness of the decay at high frequencies. 
This model is flexible enough to capture a wide range of image priors, from white noise ($p=0$) to more realistic 
natural image priors with power-law decay (e.g., $p=2$). Note that, if one includes the DC component in the loss
function, it is important to ensure that $s_0$ takes the right value. Since the mean intensity can be safely
obtained from the observed frames, we set $s_0=\langle \mathbf{I} \rangle^2$.

In the joint estimator, the hyperparameters $\vect{\beta} = \{\kappa, u_0, p, \sigma^2\}$ need to be fixed in advance and cannot 
be optimized. The noise variance $\sigma^2$can be estimated from the observed images, for instance by looking at the 
high-frequency components of the power spectrum, which are dominated by noise. However, the parameters that control 
the shape of the PSD are more difficult to set and the results depend on their choice. Since they are typically 
tuned by hand, which is a time-consuming process that requires expert knowledge and experience, it is 
customary to simply fix the value of $\kappa$ and set $p=0$ \citep[e.g.,][]{lofdahl_scharmer94}. 

In contrast, using the marginal estimator, the hyperparameters of the PSD and the noise variance $\sigma^2$ can be 
optimized together with the wavefront cofficients by minimizing the negative log-marginal likelihood. Given that all hyperparameters
are non-negative, we reparameterize them as the exponential of unconstrained parameters to ensure that they remain 
positive during the optimization.

\subsection{Numerical optimization}
The loss functions of the joint and marginal estimators are non-convex and have multiple local minima, 
so that the optimization is not trivial. We use the Adam \citep{adam14} first order optimization algorithm and make 
use of the automatic differentiation capabilities of the \texttt{PyTorch} library \citep{pytorch19}. 
We use a learning rate of 0.05 for the wavefront coefficients and a learning rate of 0.08 for the 
hyperparameters (we did not carry an exhaustive search for optimal values). Convergence is typically achieved 
after $\sim$400 iterations. Following the 
same approach used in \cite{2025A&A...703A.269A}, we try to avoid the local minima by a scheduling approach. We start
the optimization with a very simple wavefront expansion, in which only tip and tilt modes are included. After a 
certain number of iterations, we slowly increase the number of modes in the wavefront expansion until reaching the
desired amount. 

Even using this annealing approach, we found in practice that the hyperparameters of the PSD might diverge 
during the optimization. Of special relevance is $\kappa$, which controls the overall contrast of the final image, and 
can easily diverge to very large values when the number of modes is large. We think this divergence is a consequence
of the fact that, when the complexity of the PSFs increases, a huge $\kappa$ strongly decreases the  
regularization in the loss function ($s_u \gg \sigma^2$), allowing high-frequency noise to be interpreted as the effect of
very complex PSFs. This causes extreme contrast and artifacts. We will show examples of this effect in the
next section. To avoid this problem, we use a simple annealing approach in which the hyperparameters of the PSD 
are optimized only during the first iterations of the optimization, when 
the number of modes is small (typically until 5 KL modes are included) and the solution is more stable. They 
are then kept fixed for the rest of the optimization.

\section{Results}
\label{sec:results}
\subsection{Observations and numerical details}
We present the outcomes for both marginal and joint estimators by utilizing data gathered
from the CRISP and HiFI instruments. The first CRISP dataset covers a $20''\times 20''$
field of view (FoV) centered on the active region AR12767. Recorded on July 27, 2020, at
heliocentric positions $(-20'', -416'')$, this dataset was previously outlined in
\cite{2022ApJ...928..101V}. For these observations, the CRISP narrow-band (NB) channel was
positioned 65 m\AA\ to the red of the Ca \textsc{ii} 8542 \AA\ line core, employing a FWHM of
105 m\AA. The wide-band (WB) channel lies in the surrounding continuum. The spatial sampling is 0.059 arcsec/pixel. Atmospheric conditions were
moderate, with $r_0$ values ranging from 9 cm to nearly 16 cm (recorded at 500 nm via the
AO sensor) over the 20 s acquisition, averaging $\sim$12 cm for this specific data
slice. The second CRISP dataset features quiet Sun data taken at disk center on August 1,
2019, with a 31 s temporal cadence. Similar to the first set, the NB image was captured at
an offset of 65 m\AA\ to the red of the line core, while the WB channel is located in the
surrounding continuum. These data comprise a portion of the
observations detailed by \citet{2023A&A...672A.141E}. Seeing conditions were excellent,
maintaining $r_0$ measurements between 15 and 20 cm during the sequence. Finally, we
include G-band data from the High-resolution Fast Imager \citep[HiFI,][]{hifi}, situated
on the 1.5-m GREGOR solar telescope \citep{GREGOR, kleint2020}. This dataset focuses on 
AR13378, observed on July 21, 2023. The observation sequence involved capturing bursts of
500 frames every 11 s with a scale of 0.02761 arcsec/pixel. Processing and calibration were handled by the sTools pipeline
\citep{stools}, which involved a selection process where only the 50 highest-quality
frames from each burst were retained for analysis.

All results presented in this work have been obtained using an updated 
version of the \torchmfbd\ code \citep{2025A&A...703A.269A}.
The PSFs are parameterized using the KL expansion of the wavefront. In order to deal with 
anisoplanatic effects, the field of view is divided into isoplanatic patches, each of which is reconstructed
independently and merged together at the end. For the case of the CRISP observations, we use
patches of $88 \times 88$ pixels, while we use patches of $96 \times 96$ pixels for the HiFI observations. 
The reconstructions are obtained with no regularization applied to the PSFs, i.e., $J(\mathbf{H}|\vect{\beta})=0$.

\begin{figure}
    \centerline{\includegraphics[width=0.9\textwidth]{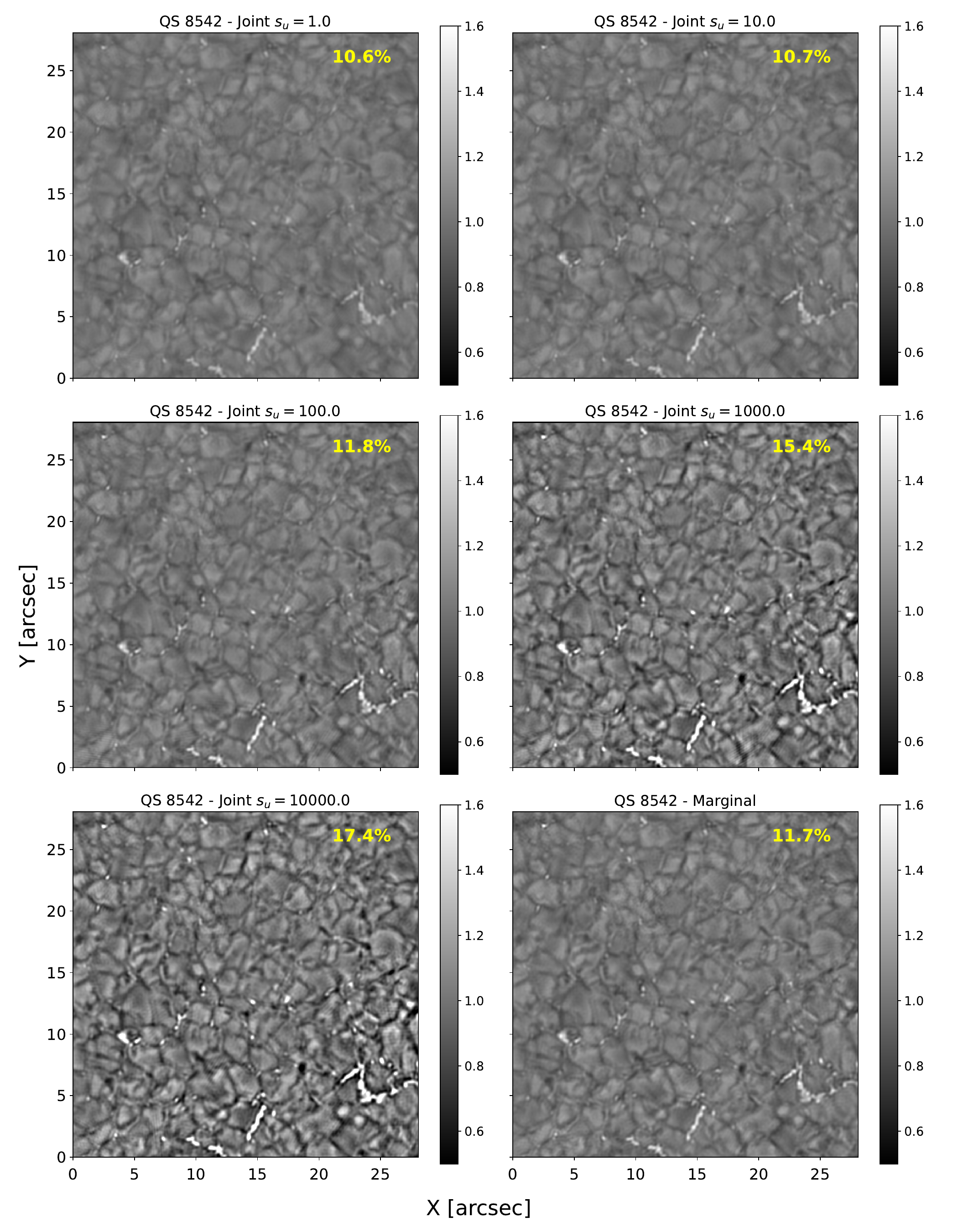}}
    \caption{Reconstruction of the quiet Sun CRISP dataset for different values of the hyperparameter 
    $K$ controlling the PSD of the object prior. The contrast of the reconstructions is very sensitive to the choice of $K$.}
    \label{fig:su}
\end{figure}

\begin{figure}
    \centerline{\includegraphics[width=\textwidth]{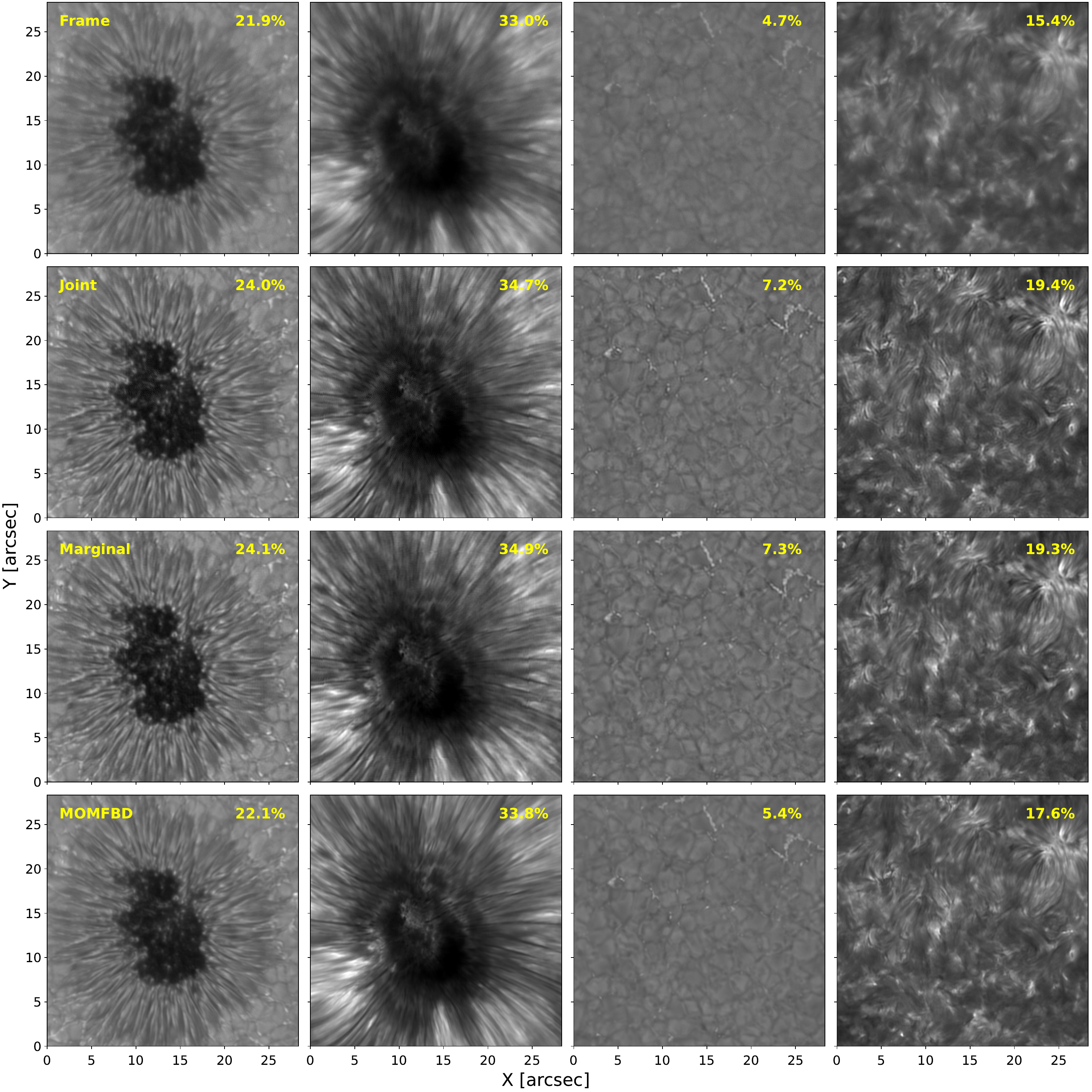}}    
    \caption{Comparison of the joint and marginal estimators obtained with \torchmfbd\ with the
    results of \momfbd.}
    \label{fig:images_8542}
\end{figure}

\subsection{Contrast of the reconstructions}
As already mentioned, the PSD of the object prior, $s_u$, plays a crucial
role in the performance of the blind deconvolution algorithm. The ratio $\sigma^2/s_u$ can be interpreted 
as an inverse signal-to-noise ratio (SNR). The smaller the value of this ratio (the larger $s_u$), the more the algorithm will 
rely on the observed data. The tuning of $s_u$ faces the fundamental problem that the spatial frequencies 
of the reconstructions are not very sensitive to its specific value once it is larger than 
a certain threshold. However, the contrast of the reconstructions is strongly dependent on
its value. This is demonstrated in Fig. \ref{fig:su}, which displays the reconstructions of the quiet Sun CRISP dataset 
with the joint estimator using $s_u=\kappa N$ ($p=0$ and $u_0 \to \infty$). For this case we use 
wavefronts with 54 KL modes. When $\kappa$ is too small, the contrast is reduced, eventually leading to a blurry reconstruction
because the algorithm is putting more strength on the prior than on the data. On the other hand, when $\kappa$ is too large, the contrast is 
artificially enhanced, leading to the appearance of very bright and very dark pixels whose contrast can be increased
with almost no control by further increasing $\kappa$. The last panel displays the reconstruction obtained with the marginal estimator, 
which gives a reconstruction more in the line of what we expect for the quiet Sun without the need of tuning
any hyperparameter.

\begin{figure}
    \centerline{\includegraphics[width=0.9\textwidth]{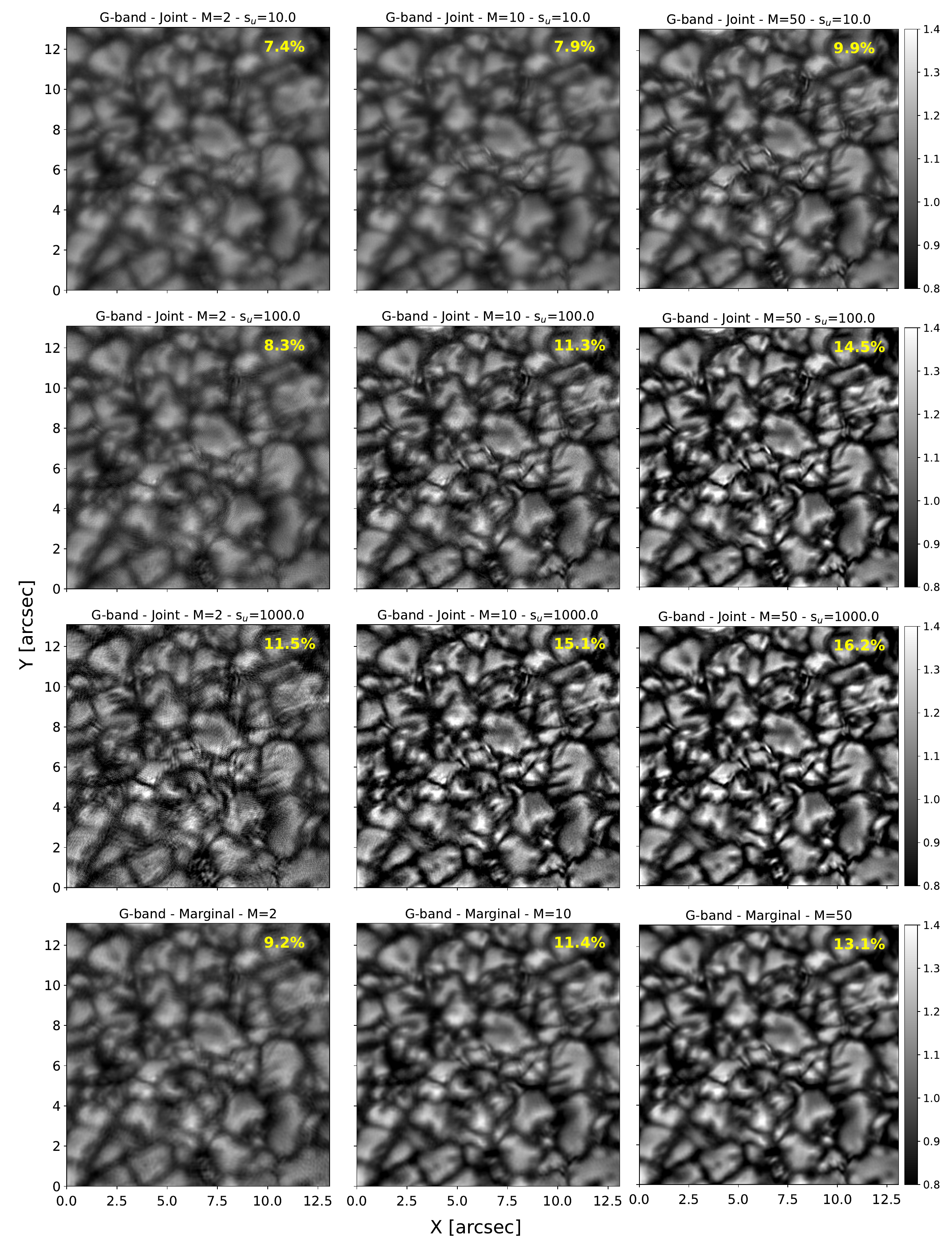}}
    \caption{Reconstruction of the quiet Sun HiFI dataset for different number of frames $M$ and different
    values of the constant PSD $s_u$. The last row displays the reconstruction obtained with the marginal estimator.}
    \label{fig:nimages}
\end{figure}

\subsection{Comparison with \momfbd}
Figure \ref{fig:images_8542} compares the results obtained with \torchmfbd\ with the results of \momfbd\
for the two CRISP datasets considered in this work. Both the WB and NB channels are used during the reconstructions as
two different objects. The upper row shows the first frame of the burst of 12 images, the second
and third rows display the joint and marginal estimators, respectively, while the last row shows the results obtained
with \momfbd. The contrast quoted in each panel has been obtained in the whole image. The joint reconstruction is computed
for $s_u=100$, while the \momfbd\ reconstructions are obtained with the default settings of the code.
In terms of visual appearance, the joint and marginal reconstructions appear to display slightly more defined
structures than the \momfbd\ results. They are more conspicuosly seen in the penumbra. Concerning the contrasts,
similar values are reached, with the marginal estimator giving slightly higher values than the joint estimator, and
both being larger than the contrast obtained with \momfbd.

\subsection{Number of images}
A similar behavior is found when the reconstructions are done with different number of frames in the burst. For
this purpose, we focus on the G-band HiFI dataset, which has a large number of frames per burst. The results
of the joint reconstruction with different values of $s_u$ and different number of frames $M$ are displayed
in the first three rows of Fig. \ref{fig:nimages}. When using only two images in the burst ($M=2$), the reconstructions 
are blurry and the contrast is similar to that of the original frames. Increasing $s_u$ slightly increases the contrast but also 
produces more artifacts in the reconstructions. The solution with $s_u=1000$ contains lots of high frequency artifacts, 
because the Wiener regularization tends to zero. The marginal reconstruction (lower row) gives no artifacts and 
automatically produces a contrast and a visual appearance that is similar to the one obtained with the joint estimator 
with $M=10$ and a small value of $s_u$. As $M$ increases, the contrast of the reconstructions increases as well, but it starts to
become out of control when $s_u$ is too large in the joint reconstructions. Again, we find that the marginal estimator
with $M=10$ produces an image with a contrast and visual appearancesimilar to the one when $M=50$ and a small 
value of $s_u$ in the joint estimator. When $M=50$, the marginal reconstruction gives a contrast that is 
slightly below that obtained with the joint estimator with $s_u=100$. Since this stability in the 
reconstructions is found automatically, the marginal reconstructor can be deployed in a more ``plug-and-play'' fashion,
without human intervention.

\begin{figure}
    \centerline{\includegraphics[width=\textwidth]{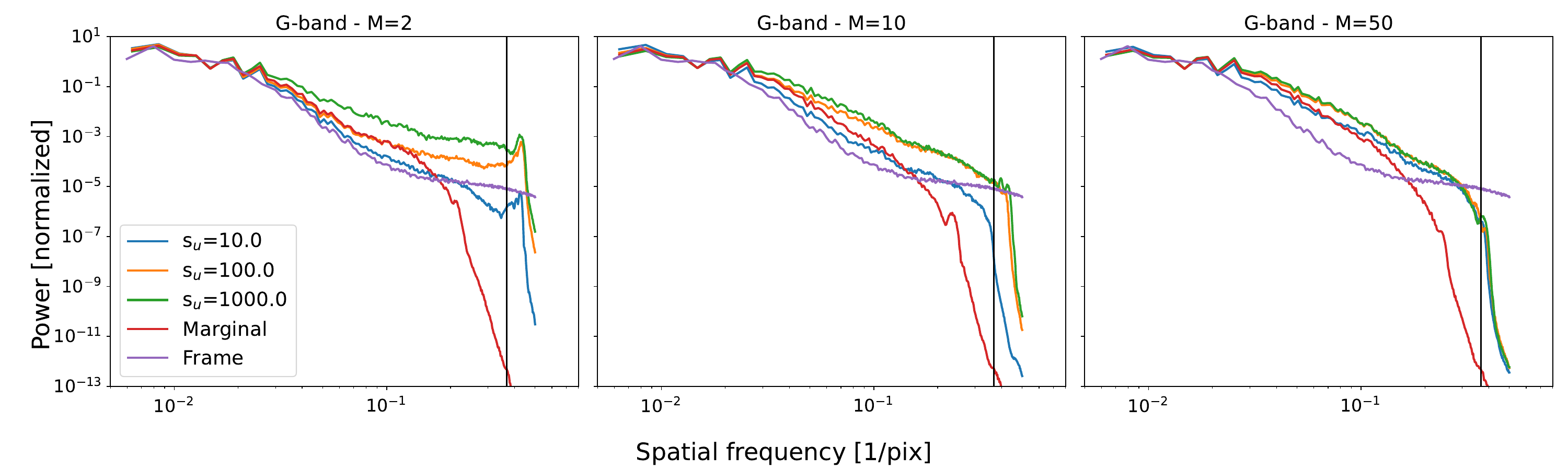}}
    \centerline{\includegraphics[width=\textwidth]{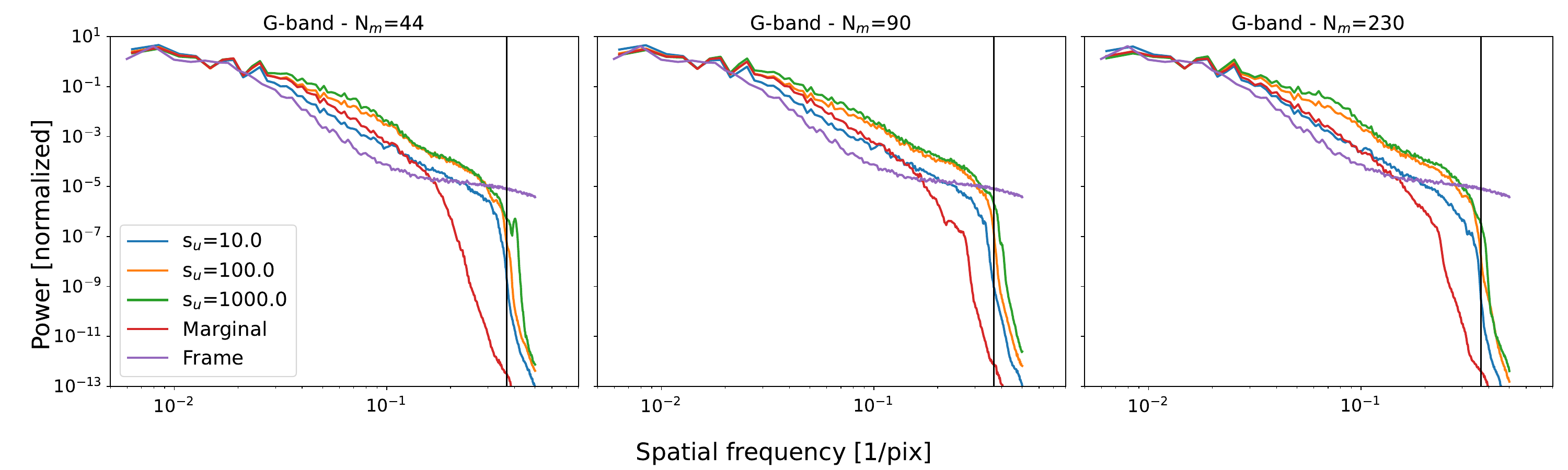}}
    \caption{Azimuthally averaged power spectra of the reconstructions for different number of images and 
    modes, correspoding to the results shown in Fig. \ref{fig:nimages} and \ref{fig:nmodes}, respectively.}
    \label{fig:power_spectra}
\end{figure}

On a more quantitative way, the influence of $M$ and $s_u$ on the reconstructions can be evaluated by looking 
at the azimuthally averaged power spectra of the reconstructions, which are shown in the upper panel of Fig. \ref{fig:power_spectra} for the different 
reconstructions shown in Fig. \ref{fig:nimages}. The vertical line indicates the diffraction limit.
The power spectrum of the original images (labeled as ``Frame'' in the figure) saturate
at high frequencies due to the noise. This is compensated for by all the reconstructions, specially
at intermediate frequencies above $\sim$0.03 pix$^{-1}$ ($\sim$0.92 arcsec). In the case of $M=2$, the reconstruction with $s_u=1000$
produces an excess of power at these intermediate frequencies, while the one with $s_u=10$ barely introduces extra power, leaving
an image that resembles the original frames. The marginal reconstruction seamlessly follows the reconstruction with $s_u=100$ but 
produces a better filtering close to the diffraction limit. As already discussed in \cite{2025A&A...703A.269A}, the power above the 
diffraction limit is an artifact produced by the patching of the FoV and its final merge. Finally, when using $M=50$,
all reconstructions produce a similar power spectrum, with less artifacts close to the diffraction limit.

\begin{figure}
    \centerline{\includegraphics[width=0.9\textwidth]{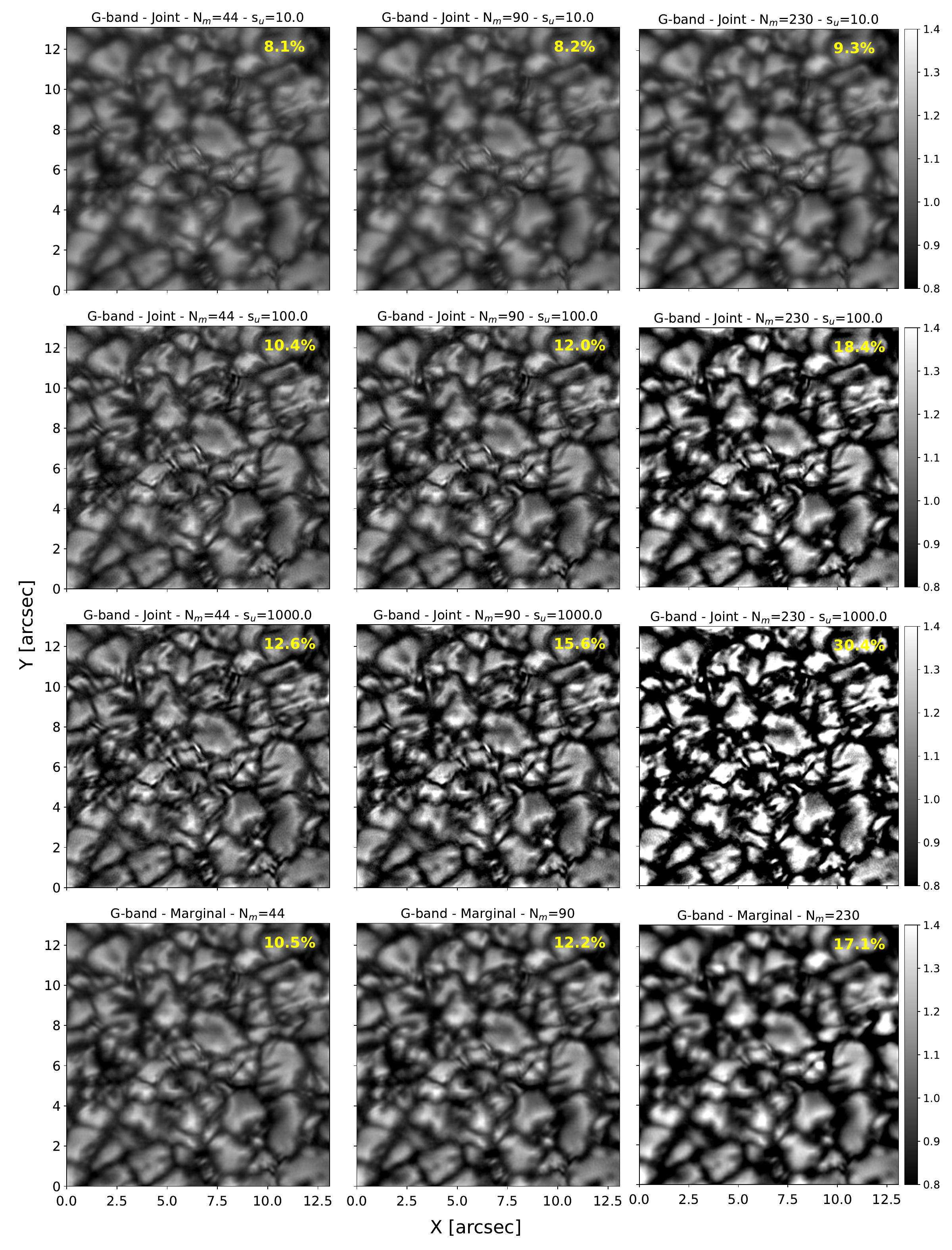}}
    \caption{Reconstruction of the quiet Sun HiFI dataset for different number of modes $N$ and different
    values of the constant PSD $s_u$. The last row displays the reconstruction obtained with the marginal estimator.}
    \label{fig:nmodes}
\end{figure}

\begin{figure}
    \centerline{\includegraphics[width=\textwidth]{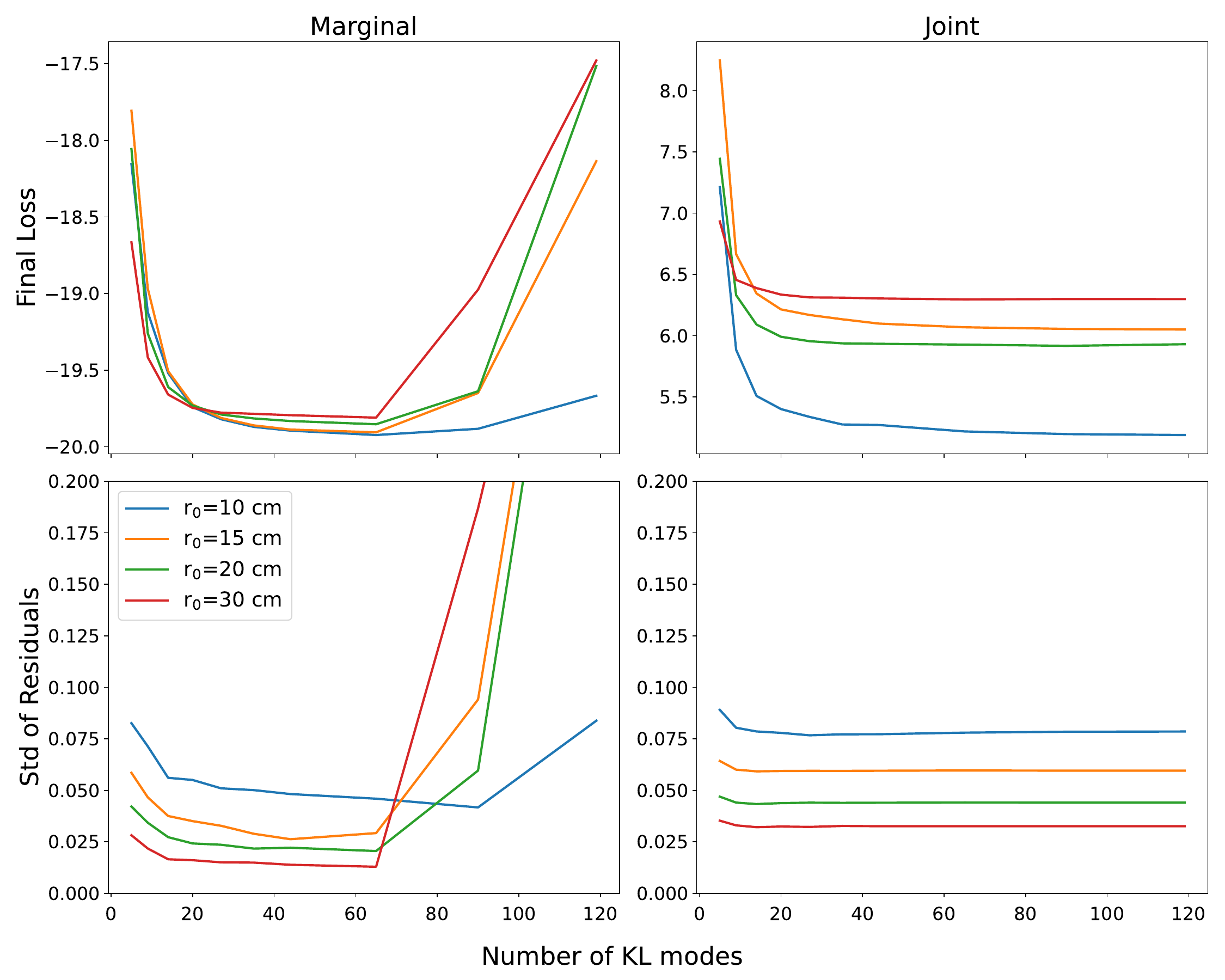}}
    \caption{Loss and standard deviation of the residuals as a function of the number of modes.}
    \label{fig:optimal_modes}
\end{figure}

\subsection{Number of modes}
The number of modes used to parameterize the PSFs also has a strong influence on the reconstructions
and it is something that is normally tuned by hand. When too few modes are used, the reconstructions are blurry 
and the contrast is low. As the number of modes increases, the reconstructions can potentially improve
if enough information is present in the observed frames. However, the higher order modes are less certain and
the contrast becomes out of control if the PSD of the object prior is not properly tuned. This is 
shown in Fig. \ref{fig:nmodes}, where we display the reconstructions of the quiet Sun HiFI dataset with 
different number of modes and different values of $s_u$. In fact, when the number of modes is increased,
the value of $s_u$ needs to be decreased to avoid the appearance of artifacts in the reconstructions. 
The marginal reconstruction (lower row) gives an automatically
controlled contrast, which never becomes out of control even if a large number of modes is used.

In terms of the power spectrum (see the lower panels of Fig. \ref{fig:power_spectra}), we find that the
joint reconstruction with $s_u=10$ is not able to add enough frequencies in the intermediate regime for
all cases. The marginal reconstruction seamlessly follows the joint reconstruction between $s_u=10$ and $s_u=100$, but it 
produces a better filtering close to the diffraction limit. The power spectra of all reconstructions 
when the number of modes is too large are very similar, and the only impact lies in the increased
contrast. In other words, no new spatial structures appear but the contrast of the existing structures is 
increased.

The regularization provided by the log-determinant term in the marginal estimator can be also 
used to automatically select the optimal number of modes to be used in the reconstruction. This is done by 
looking at the value of the loss function as a function of the number of modes. The optimal number of modes is 
the one that minimizes the loss function, which is a balance between the data fit and the regularization 
provided by the log-determinant term. To have full control over the results and verify the behavior of the
method with the strength of the turbulence, we carry out a simple experiment with synthetic data. We use a 
synthetic intensity map in the continuum around the 6301-6302 \AA\ Fe \textsc{i} doublet. This map was already used 
by \cite{2023A&A...675A..93Q} and the snaphot has been obtained from a magneto-hydrodynamic simulation of 
the solar atmosphere with the \texttt{Mancha3D} code \citep{2024SoPh..299...23M}.
We then generate a burst of 10 images by 
convolving the synthetic image with 10 different PSFs with Kolmogorov turbulence with different values 
of the Fried radius ($r_0$). We assume a maximum of 210 KL modes for the turbulent wavefronts. We also
add Gaussian noise with $\sigma=0.01$ in units of the continuum intensity.

We carry out the reconstructions with different number of modes using both the joint (with a fixed value of $s_u=100$)
and marginal estimators and look at the value of the loss function as a function of the number of modes. The results are shown in 
Fig. \ref{fig:optimal_modes}. The loss functions (shown in the upper panels) of the marginal estimator have 
a clear minimum, while the joint estimator decreases monotonically as the number of modes increases. The number of modes required for the 
optimal reconstruction is larger for the case with stronger turbulence (smaller $r_0$), which is expected 
since more modes are needed to capture the complexity of the PSFs. Of the maximum number of 210 modes, the marginal 
estimator points to an optimal number of modes in the range between 60 and 80 for the case with $r_0=10$ cm, and 
less than 60 KL modes for the case with $r_0=30$ cm. The lower panel of Fig. \ref{fig:optimal_modes} shows the 
standard deviation of the residuals between the reconstructed images and the original image as a function of the number 
of modes. This residual, which is a measure of the quality of the reconstruction, strongly correlates with the
loss function. Therefore, we conclude that optimizing the marginal loss function is a good strategy to automatically 
select the optimal number of modes to be used in the reconstruction.

As a subproduct, the adaptive character of the marginal estimator allows the residual to be smaller than
the one obtained with the joint estimator with a fixed value of $s_u$ for all values of $r_0$. It is, in principle, potentially possible
to optimize by trial-and-error the optimal value of $s_u$ for the joint estimator to get a similar performance, 
but this is a time-consuming process that requires expert knowledge and experience. 

\section{Conclusions and future work}
This work introduces a marginal estimator for the multi-object multi-frame blind
deconvolution problem, addressing the inherent limitations of traditional joint
Maximum Likelihood estimation in solar imaging. By employing a Bayesian
framework to marginalize over the observed objects, the proposed method provides
a stable reconstruction that is potentially less sensitive to high-frequency noise.
We find that the marginal estimator offers several advantages over the joint estimator.
The first is an enhanced regularization, a consequence of the marginalization, which 
naturally accounts for object uncertainty, preventing the reconstruction
algorithm from assigning noise power to high-order aberrations.
The second is a more robust contrast control, as the marginal estimator is almost insensitive 
to the choice of the hyperparameters that control the PSD of the object prior. Since these hyperparameters
can be optimized, this allows for a more ``plug-and-play'' deployment of the algorithm, without the need for expert tuning. 
Finally, the method is easy to implement, since the merit function to be optimized
contains only an additional log-determinant term with respect to the joint estimator.
The necessary modifications to existing blind deconvolution pipelines are minimal. 
The marginal estimator has been already integrated into the open-source \torchmfbd\ package, facilitating its use in
the solar physics community.

Finally, the marginal estimator opens up the door to a fully automated MOMFBD code, in which the number of 
modes used to parameterize the PSFs is automatically selected. Directly adding the number of modes as a parameter 
during the optimization is potentially possible, although difficult because of its discrete nature. One cannot 
directly use gradient-based optimization methods like the ones used in this
work. We are exploring several strategies to implement in \texttt{torchmfbd}. Among the options, we will explore
the use of a Gumbel-Softmax distribution \citep{2016arXiv161101144J} to relax the discrete nature of the number of modes or
the use of a continuum parametric sigmoid mask that returns the importance of each mode.

%%%%%%%%%%%%%%%%%%%%%%%%%%%%%%%%%%%%%%%%%%%%%%%%%%%%%%%%%%%%%%%%%%%%%%%%%%%
\begin{acks}
I thank C. Quintero Noda for providing the synthetic snapshot used in the experiment of Sect. \ref{sec:results}
and M. L\"ofdahl for a careful reading of the manuscript and suggesting the possibility of inferring the optimal
number of modes. I acknowledge the 
community effort devoted to the development of the following open-source packages that were
used in this work: \texttt{numpy} \citep[\texttt{numpy.org},][]{numpy20}, 
\texttt{matplotlib} \citep[\texttt{matplotlib.org},][]{matplotlib}, and \texttt{PyTorch} 
\citep[\texttt{pytorch.org},][]{pytorch19}.
\end{acks}

\begin{authorcontribution}
All the work was carried out by A.A.R.
\end{authorcontribution}

\begin{fundinginformation}
I acknowledge funding from the Agencia Estatal de Investigación del Ministerio de Ciencia, Innovaci\'on y 
Universidades (MCIU/AEI) under grant ``Polarimetric Inference of Magnetic Fields'' and the European Regional 
Development Fund (ERDF) with reference PID2022-136563NB-I00/10.13039/501100011033.
\end{fundinginformation}

\begin{codeavailability}
The code, together with scripts for reproducing the results of this paper, is available at 
\url{https://github.com/aasensio/torchmfbd}.
\end{codeavailability}

\begin{ethics}
\begin{conflict}
The authors declare that they have no conflicts of interest.
\end{conflict}
\end{ethics}

\bibliographystyle{spr-mp-sola}
     % name your Bibtex file containing your references (.bib)
\bibliography{biblio}

\end{document}